# Fiber Nonlinearity Compensation of Coherent Signals Using Deep Photonic Reservoir Computer

Yi-Wei Shen,[1,2] Rui-Qian Li,[1,2] Zheng-Can Sun,[1] Xing Li,[3,a)] Xinyu Liu,[4] Shanshan Yu,[4] and Cheng Wang[1,5,a)]

## AFFILIATIONS

[1]School of Information Science and Technology, ShanghaiTech University, Shanghai 201210, China
[2]Guangjiu Technology, Shanghai 201210, China
[3]State Key Laboratory of Photonics and Communications, Intelligent Microwave Lightwave Integration Innovation Center (imLic), Department of Electronic Engineering, Shanghai Jiao Tong University, Shanghai 200240, China
[4]Xizhi Technology Co., Ltd, Shanghai 201210, China
[5]State Key Laboratory of Quantum Functional Materials, ShanghaiTech University, Shanghai 201210, China

[a)]**Authors to whom correspondence should be addressed:** wangcheng1@shanghaitech.edu.cn and lixing85715@sjtu.edu.cn

## ABSTRACT

Photonic reservoir computer (PRC) is a promising optical computing framework for high-speed optical signal processing, and various reports have shown its functionality of linear equalization for intensity-modulation direct-detection communication links. However, coherent communication links suffer more from nonlinear impairment, whereas its nonlinear equalization is very challenging. Here we demonstrate the nonlinear equalization of coherent 16-level quadrature amplitude modulation (16-QAM) signals using a deep PRC in experiment. The deep PRC consists of cascading injection-locked semiconductor lasers with optical feedback loops. For 16-QAM signals with a transmission rate of 240 Gbps and a launch power of 12 dBm, the single-channel PRC with 3 hidden layers raises the Q factor by as high as 0.58 dB, while the dual-channel PRC with 2 hidden layers raises the Q factor by 0.55 dB. In addition, we prove that the deep PRC is able to equalize 16-QAM signals of different launch powers and transmission distances.

## I. INTRODUCTION

In optical fiber communication links, optical signals suffer from both linear distortion and nonlinear distortion. The linear distortion primarily arises from the chromatic dispersion effect, which has been well compensated for by digital signal processing (DSP) algorithms like feedforward equalizer (FFE) and decision feedback equalizer.[1-3] On the other hand, the nonlinear distortion mainly originates from the Kerr nonlinearity. Common nonlinear impairment compensation algorithms include digital backpropagation (DBP),[4] perturbation-based approach,[5] Volterra series, and so on.[6] However, these nonlinear algorithms are computationally intensive and hence consume too high power for practical DSP implementations. In order to reduce the power consumption, various deep-learning-based nonlinear equalizers have been proposed.[7-12] However, there remains tradeoff between the complexity of neural networks and the performance of nonlinear equalization.

In recent years, researchers have proposed utilizing photonic reservoir computers (PRCs) for the fiber impairment compensation of optical signals.[13-20] PRCs are analog recurrent neural networks, which are implemented by various photonic hardware, such as semiconductor lasers,[21] multimode interferometers,[22] modulators,[23] and microrings.[24] In particular, only the output layer of PRCs requires training through the linear regression algorithm, while the weights in the input layer and in the hidden reservoir layers are randomly fixed. As a result, PRCs gain the capability of real-time training, whenever the weights need to be adapted to altered operating conditions or changed tasks. There have been extensive reports on the linear and nonlinear equalizations of intensity-modulation direct-detection (IMDD) optical signals using PRCs in literature.[13-16] In comparison with IMDD communication systems, coherent communication systems suffer from much stronger nonlinear impairment because both the amplitude and phase of the electric field are multiplexed.[25,26] In simulation, Masaad *et al.* utilized the multimode-interferometer (MMI)-based PRC in combination with a Kramers–Kronig receiver to compensate for the nonlinearity of 64-level quadrature-amplitude modulation (QAM) signals. It was shown that this scheme surpassed the FFE algorithm in processing the 64-QAM signal at a rate of 64-GBaud over 100 km.[17] Zelaci *et al.* used the MMI-based PRC to reduce the carrier-to-signal power ratio requirement of a Kramers–Kronig receiver, when equalizing a 32-GBaud 16-QAM signal over 80 km.[18] Zhang *et al.* found that the laser-based PRC outperformed the digital DBP (three steps per span) in the nonlinear equalization of 16-QAM signals, while the modulator-based PRC could only compensate for the linear chromatic dispersion.[19] In experiment, Masaad *et al*. successfully demonstrated the equalization of 28-GBaud 16-QAM signals over 20 km, using the MMI-based PRC.[20]

Despite the above advances, a key constraint of most reported PRC systems for nonlinearity equalization is the shallow architecture, which consists of only one hidden reservoir layer. Inspired from deep learning theory, the deep reservoir structure substantially improves the representational capacity compared with the shallow one.[27,28] In our previous work, we introduced a deep PRC composed of cascading injection-locked semiconductor lasers, which connects successive reservoir layers in an optical approach.[29,30] In each hidden layer, a large number of interconnected optical neurons are generated from the nonlinear dynamics of the semiconductor laser with optical feedback.[31,32] Recently, we theoretically applied the deep PRC in the nonlinearity equalization of 16-QAM signals, which proved that the Q factor was raised by more than 1 dB for launch powers above 10 dBm at a data rate of 30 GBaud over 50-km transmission.[33]

Following our previous theoretical analysis in Ref. 33, this work experimentally demonstrates the deep PRC designed for the nonlinear equalization of 16-QAM signals. The deep PRC system employs cascading injection-locked distributed feedback (DFB) lasers with optical feedback, so as to construct successively connected hidden reservoir layers. For the nonlinear impairment compensation of 30-GBaud 16-QAM signals over a transmission distance of 50 km, the single-channel PRC with 3 hidden layers successfully improves the Q factor by 0.58 dB at a launch power of 12 dBm, which yields a bit error rate (BER) reduction of 44%, compared to the DSP module without nonlinear equalization. Meanwhile, the dual-channel PRC with 2 hidden layers raises the Q factor by 0.55 dB, corresponding to a BER reduction of 42%. In addition, we prove that the deep PRC is able to deal with 16-QAM signals of different launch powers and transmission distances.

## II. FRAMEWORK AND EXPERIMENTAL SETUP

Figure 1 illustrates the framework of the transmission system delivering the dual-polarization 16-QAM signal, the DSP module for conventional signal processing, and the deep PRC module for the nonlinearity equalization. At the transmitter, both in-phase (I) and quadrature (Q) components of the 16-QAM signal are superimposed onto the carrier wave of the laser through the modulator. The modulated signal travels through a span of optical fiber, and is amplified by an erbium-doped fiber amplifier (EDFA) before reaching the receiver. The receiver consists of a photodetector and an analog-to-digital converter (ADC). The DSP module executes multiple functions in series, including chromatic dispersion compensation (CDC),[34] frequency offset estimation (FOE),[35] polarization demultiplexing, and carrier phase estimation (CPE).[36-38] The DSP output provides the demodulated I and Q components for both x polarization and y polarization. The above transmission link is simulated by an open-source platform coined Intelligent Fiber Transmission Simulation (IFTS) Python tool.[39] The output signal from the DSP is fed into the deep PRC module in the experiment, which is dedicated to compensating for the nonlinear fiber impairment. In this work, the deep PRC is applied only to the x-polarization component, while the y-polarization component can be equalized using the same setup or using the wavelength division multiplexing (WDM) technique discussed in Ref. 33. In the simulation of the optical transmission system, the dual-polarization 16-QAM signal is modulated at a symbol rate of 30 GBaud, yielding a bit rate of 240 Gbps. The default launch power at the transmitter is 12 dBm and the default transmission distance

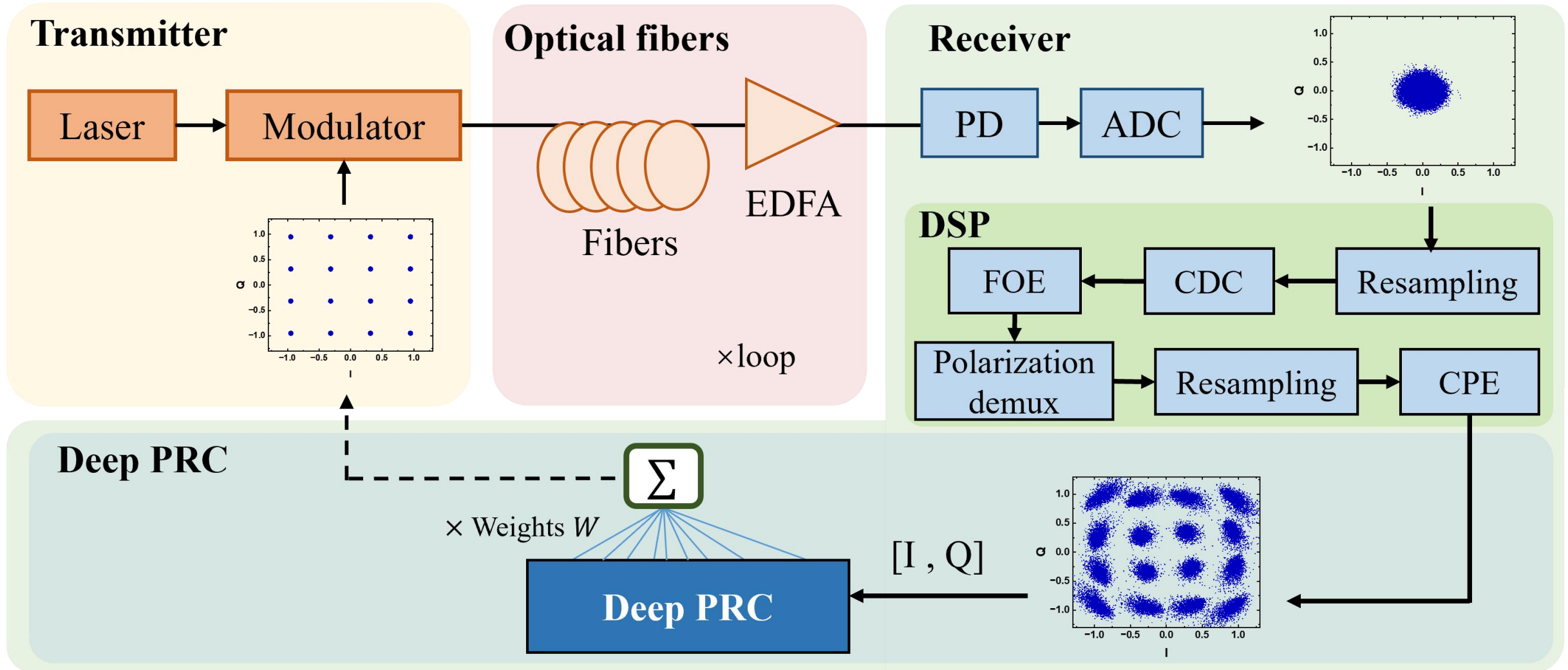

**FIG. 1.** Pipeline of the signal transmission and signal processing with deep PRC. EDFA: erbium-doped fiber amplifier; CDC: chromatic dispersion compensation; FOE: frequency offset estimation; CPE: carrier phase estimation.

Is 50 km, unless stated otherwise. In the simulation, the carrier wavelength is 1550 nm, the fiber loss coefficient is 0.2 dB/km, the dispersion parameter is 16.75 ps/nm/km, the fiber nonlinear coefficient is 1.32 $W^{-1}$ $km^{-1}$, the EDFA noise figure is 5 dB, and the polarization mode dispersion is neglected. Figure 2 shows representative constellations of the 16-QAM signal at the transmitter [see Fig. 2(a)] and after the DSP module [see Fig. 2(b)], respectively. It is shown that the constellation in Fig. 2(b) remains notably distorted, primarily due to the fiber nonlinearity. This distorted signal is then sent to the deep PRC for the nonlinearity equalization in the following setup.

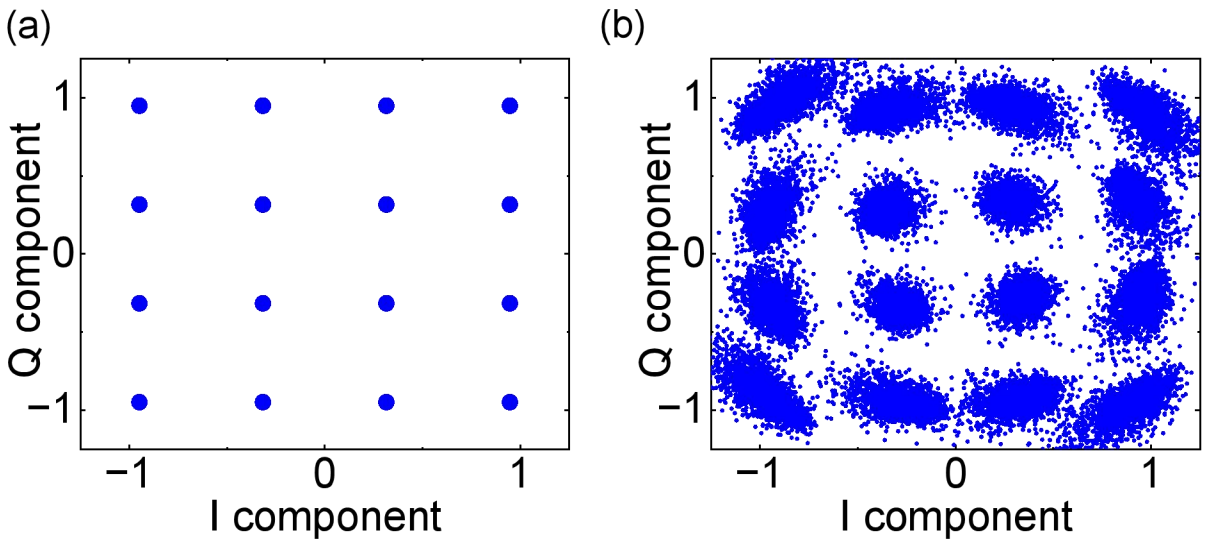

**FIG. 2.** Simulated constellation of the 16-QAM signal (a) at the transmitter and (b) after the DSP module.

Figure 3 illustrates the experimental setup of the deep PRC with four hidden layers. A tunable external-cavity laser (Santec TSL-710) serves as the master laser (ML), and its output is amplified by an EDFA. The I and Q signals output from the DSP are multiplied by a 4-level random mask consisting of values {-1, -1/3, 1/3, 1}. In comparison with the commonly used binary mask, this 4-level mask helps to enrich the neuron states.[41] These masked signals are generated by an arbitrary waveform generator (AWG, Keysight, 25 GHz bandwidth) and subsequently superimposed onto the ML carrier wave, using a Mach-Zehnder intensity modulator (EOSPACE, 40 GHz bandwidth). The polarization of the ML output is aligned with the modulator via a polarization controller and is subsequently readjusted to match the polarization of the slave laser (SL) in the first hidden layer. All SLs are DFB lasers with a lasing wavelength around 1550 nm. In each reservoir layer, the optical feedback loop is constructed using an optical circulator and two couplers. The feedback strength is finely tuned by an optical attenuator. Following the second layer, the light is amplified by another EDFA before injected into the third one. In the readout layer, all the neuron states in the four layers are detected by broadband photodiodes (PDs) and then recorded on a high-speed digital oscilloscope (OSC, Keysight, 59 GHz bandwidth). The AWG and OSC operate at sampling rates of 60 GSa/s and 80 GSa/s, respectively. The optical spectra are monitored by an optical spectrum analyzer (Yokogawa) with a resolution of 0.02 nm.

In the experiment, all the four SLs share a common lasing threshold of $I_{th}$ = 8.0 mA. The resonance frequency of the SLs is measured to be approximately 5.4 GHz (corresponding to a characteristic time of 0.19 ns), which is extracted from the intensity noise spectrum. Unless otherwise specified, the operating parameters of the deep PRC used in the experiment are listed in Table 1. In this Table, the feedback ratio is defined as the optical power ratio of the reflected light to the emitted one. The injection ratio is defined as the ratio between the optical power from the laser in the preceding layer and the optical power of the laser inthe subsequent one. The detuning frequency refers to the lasing frequency difference between two lasers. The delay times ($\tau_1$-$\tau_4$) of the four feedback loops deviate from integer multiples of the clock cycle, so as to avoid any resonance effect.[42] All the parameters in Table 1 ensure the operation in the stable regime both for optical injection and for optical feedback. The modulation frequency of the modulator is fixed at 20 GHz, yielding a neuron interval of $\theta = 0.05$ ns. Each hidden layer is configured with $N = 40$ neurons by default, resulting in a clock cycle of $T_c = 2.0$ ns ($T_c = \theta \times N$). The deep PRC is trained using 50,000 random symbols, and its performance is evaluated on a test set of 20,000 symbols. The equalization performance is quantified by the Q factor, which is derived from the BER by $Q = 20\log_{10}(\sqrt{2} \cdot erfcinv(2 \cdot BER))$, with *erfcinv* being the inverse complementary error function.[43] A lower BER results in a higher Q factor, indicating better equalization performance.

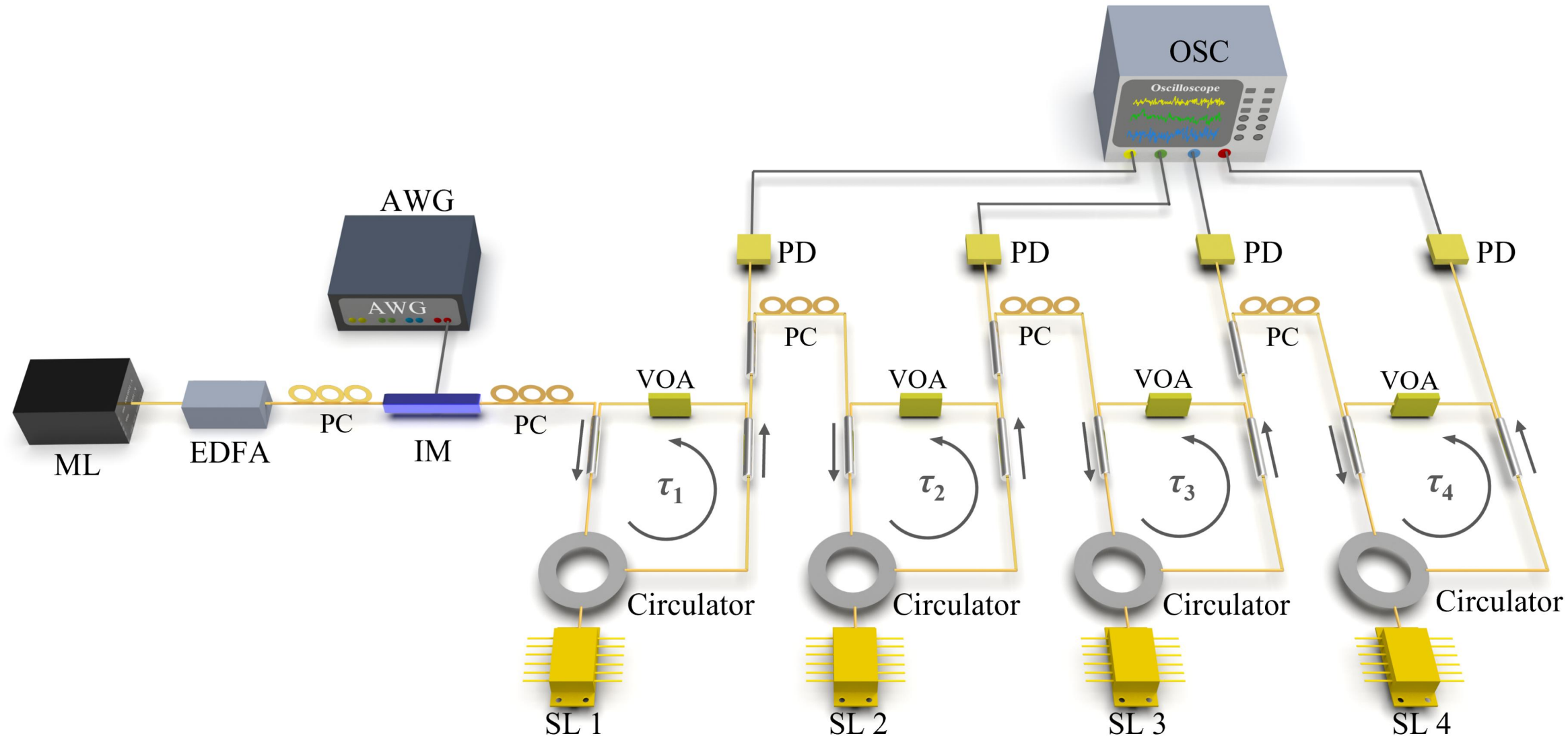

**FIG. 3.** Experimental setup of the 4-layer deep PRC. ML: master laser; EDFA: erbium-doped fiber amplifier; IM: intensity modulator; SL: slave laser; AWG: arbitrary waveform generator; OSC: oscilloscope; PD: photodiode; PC: polarization controller; VOA: variable optical attenuator.

**TABLE I.** Main parameters of the laser and the deep PRC.

| Parameter | Laser 1 | Laser 2 | Laser 3 | Laser 4 |
|---|---|---|---|---|
| Laser current | $8.0\times I_{th}$ | $2.5\times I_{th}$ | $8.0\times I_{th}$ | $2.5\times I_{th}$ |
| Laser power | 13.4 mW | 2.8 mW | 13.6 mW | 2.7 mW |
| Feedback delay | 57.0 ns | 56.9 ns | 57.2 ns | 57.2 ns |
| Feedback ratio | -30 dB | -30 dB | -30 dB | -30 dB |
| Injection ratio | 4.0 | 3.0 | 3.0 | 3.0 |
| Detuning frequency | -31.2 GHz | -3.8 GHz | -13.7 GHz | -2.0 GHz |

## III. EXPERIMENTAL RESULTS

Under the free-running condition (without optical injection and optical feedback), the lasing wavelengths of four DFB lasers in each reservoir layer in Fig. 4 are 1549.63, 1549.85, 1549.77, and 1549.86 nm (dash-dot curves), respectively. Once applying optical injection from the ML with a wavelength of 1549.88 nm, all the four SLs are phase-locked to the ML, and hence share the same wavelength (solid curves). Figure 5 shows that the DSP module alone provides a Q factor of 8.62 dB (dashed line), corresponding to a BER of $3.5 \times 10^{-3}$. When applying the deep PRC processing, the Q factor increases from 8.77 dB for a neuron number per layer of $N = 10$ up to the maximum of 9.04 dB for $N = 40$.

The performance saturates when further raising the neuron number. The PRC performance is determined not only by the readout weight, but also by the operation conditions, which act as hyperparameters of the neural network.[33] Figure 6 systematically optimizes the optical injection conditions and the optical feedback conditions layer by layer. For each layer, the parameters are tuned in the sequential order of detuning frequency, injection ratio, feedback ratio, and feedback delay. As illustrated in Fig. 6(a), through varying the detuning frequency from the saddle-node bifurcation side to the Hopf bifurcation side, the optimal Q factor is achieved at -30 GHz for the first layer, at -4 GHz for the second layer, at -13 GHz for the third layer and 0 GHz for the fourth layer, respectively. Therefore, the optimal performance of the deep PRC is generally achieved at the negative detuning side within the stable locking regime, bounded by the saddle-node bifurcation and the Hopf bifurcation. Figure 6(b) shows that a higher injection ratio improves the Q factor in every layer, which can be attributed to the improved signal-to-noise ratio (SNR) of the deep PRC system. In contrast, Fig. 6(c) and Fig. 6(d) demonstrate that the Q factor has little variation when the feedback ratio or the feedback delay time changes, provided that the optical feedback is operated in the stable regime below the critical feedback level. The critical feedback level of the SLs is measured to be approximately -20 dB, beyond which nonlinear pulsations such as chaotic oscillations emerge.[40] It is remarked that the feedback delay time in the setup is more than 15 times longer than the clock cycle ($T_c = 2.0$ ns), which is far beyond the optimal value ($1.5\times T_c$) concluded in our previous theoretical work.[33] Future work will shorten the fiber loops to further improve the deep PRC performance. Figure 6 also analyzes the contribution of every reservoir layer to the Q factor. Interestingly, it is found that the first layer contributes the most to the improvement of the Q factor, while the deeper layer generally contributes less. This is in contrast to the equalization of IMDD signals in our previous work,[30] where the deeper layer contributes more to the BER improvement. This is understandable because the performance of a certain task is not monotonically dependent on the representation richness of neural networks, although the deeper layer indeed provides richer neuron dynamics.[44] In addition, the EDFA after the second layer reduces the SNR of the third and the fourth layers, which may affect the contribution to the Q factor as well.[45,46]

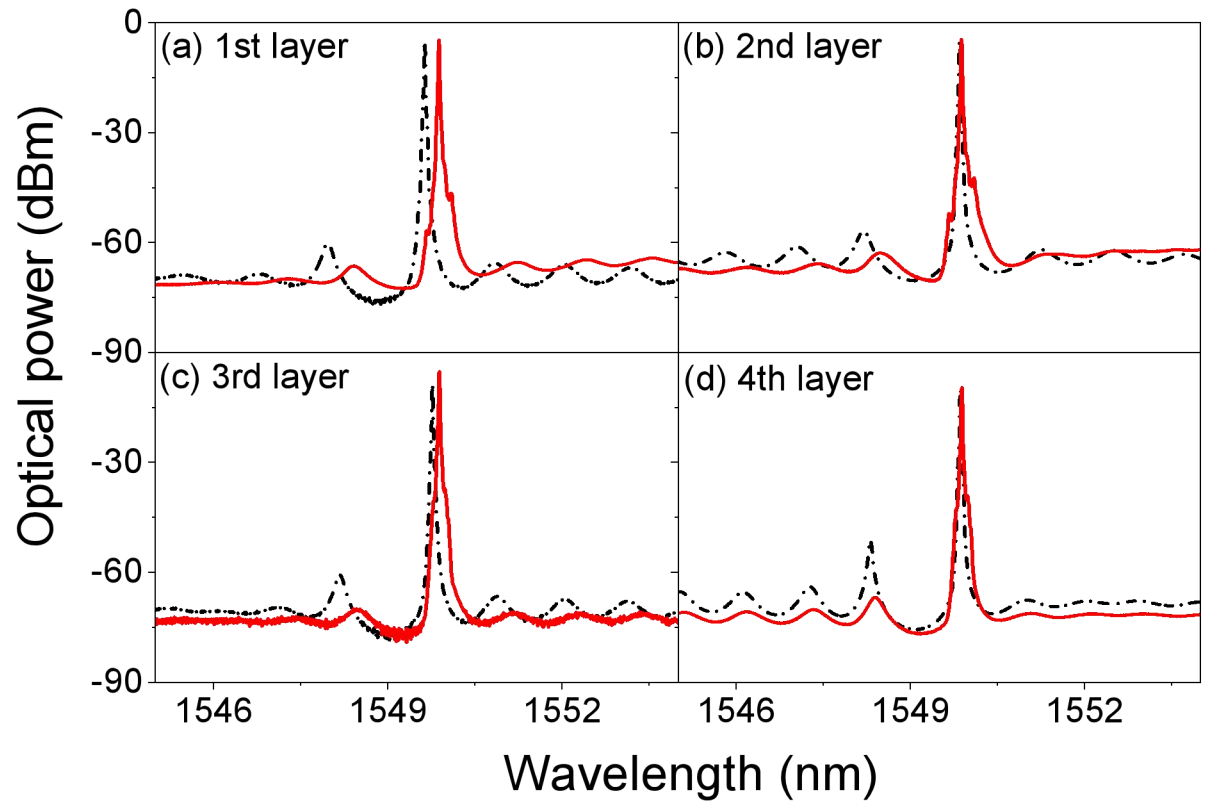

**FIG. 4.** Optical spectra of the four DFB lasers with (solid curves) and without (dash-dot curves) optical injection.

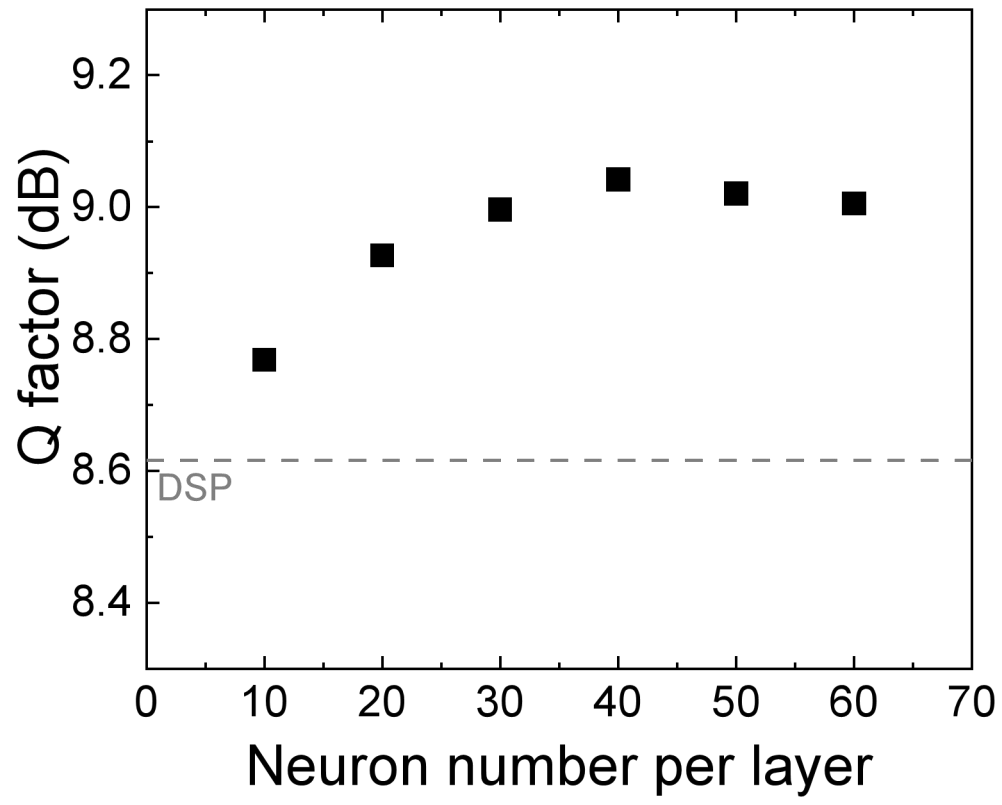

**FIG. 5.** Impact of the neuron number on the Q factor. The dashed line denotes the Q factor (8.62 dB) obtained by the DSP alone.

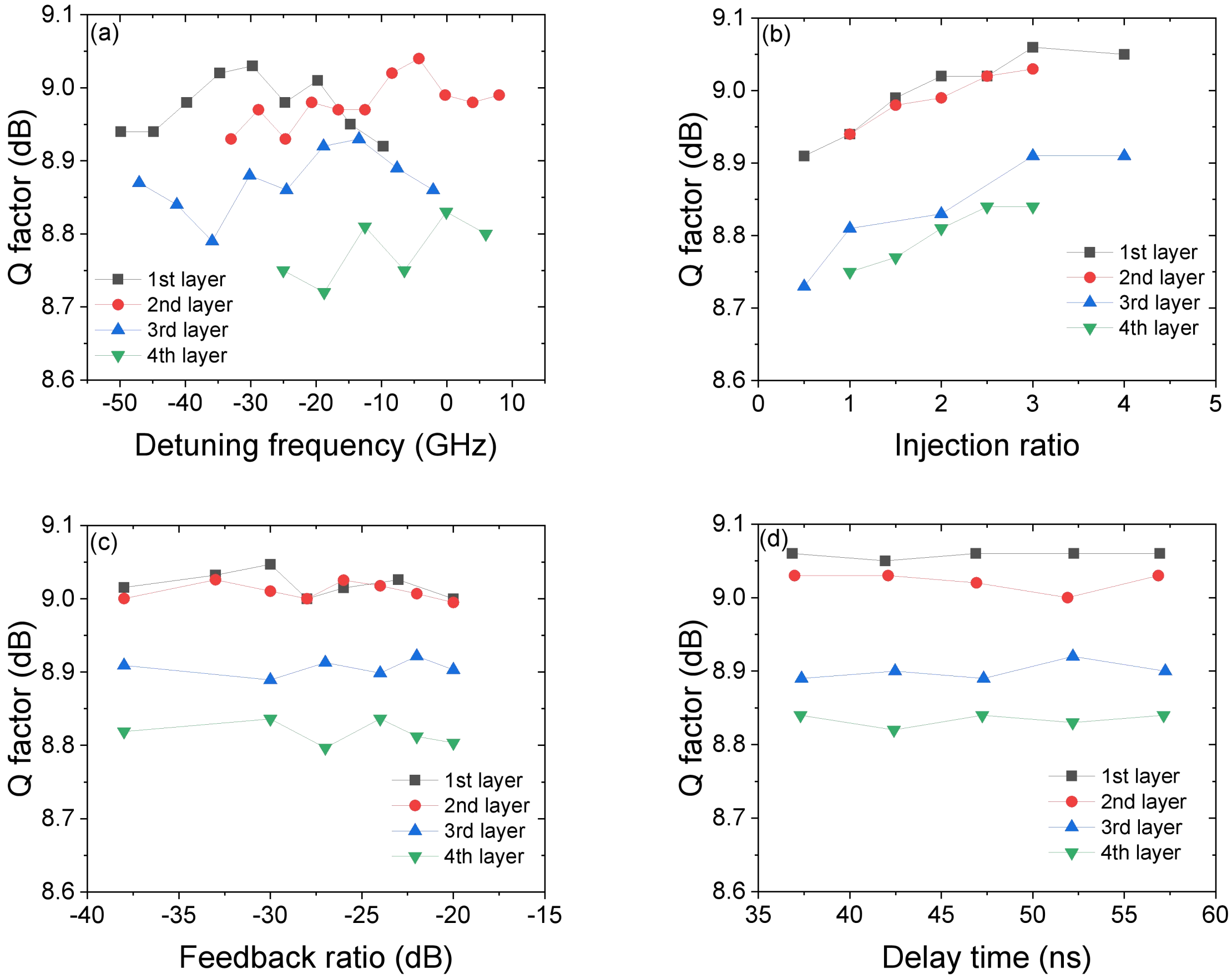

**FIG. 6.** Effects of (a) detuning frequency, (b) injection ratio, (c) feedback ratio, and (d) feedback delay time on the Q factor for each layer. The neuron number is 40 in each layer. The other parameters are listed in Table 1.

With the optimized operation parameters in Fig. 6, Fig. 7(a) shows the best achievable Q factor for different PRC depths. With one hidden layer, the PRC improves the Q factor from 8.62 dB (DSP alone, dashed line) to 9.04 dB, yielding a Q factor gain of 0.42 dB. The Q factor rises to the maximum of 9.20 dB at the depth of three, corresponding to a Q factor gain of 0.58 dB. The maximum Q factor translates to a minimum BER of $2.0 \times 10^{-3}$, which is reduced by approximately 44% in comparison with the DSP. The equalization performance saturates when the depth is further raised.[33] Therefore, although the deeper layer in Fig. 6 contributes less to the Q factor, combined neuron states of all the hidden layers help to raise the Q factor. Figure 7(b) compares the constellation diagrams of the 16-QAM signal with (red) and without (blue) the nonlinear equalization via the PRC with three hidden layers. The constellation quality is clear improved by the deep PRC, where the data points are more tightly clustered compared to that by the DSP alone.

In order to evaluate the generalization capability of the deep PRC, we investigate its equalization performance for different launch powers and different transmission distances in Fig. 8. It is stressed that the operation parameters used for the evaluation are listed in Table 1, without further optimization. As shown in Fig. 8(a), the Q factor declines with increasing launch power both for the deep PRC and for the DSP. Meanwhile, the Q factor gain generally increases from 0.31 dB (BER reduction of 39%) at 10 dBm to 0.64 dB (BER reduction of 22%) at 15 dBm, proving its strong capability in compensating for the nonlinear distortion of 16 QAM signals. This nonlinearity compensation capability arises from the rich intrinsic nonlinear dynamics of the deep PRC system. Both the Q factor and the gain at the launch power of 12 dBm deviate from overall evolution trend, because the parameters listed in Table 1 are optimized at this power. It is remarked that Fig. 8(a) does not take into account launch powers below 10 dBm, because the BER of the DSP alone falls below $1.25 \times 10^{-5}$. Such a low BER is beyond the limit of the test set of 20,000 symbols. Figure 8(b) presents that the Q factors of both the DSP and the deep

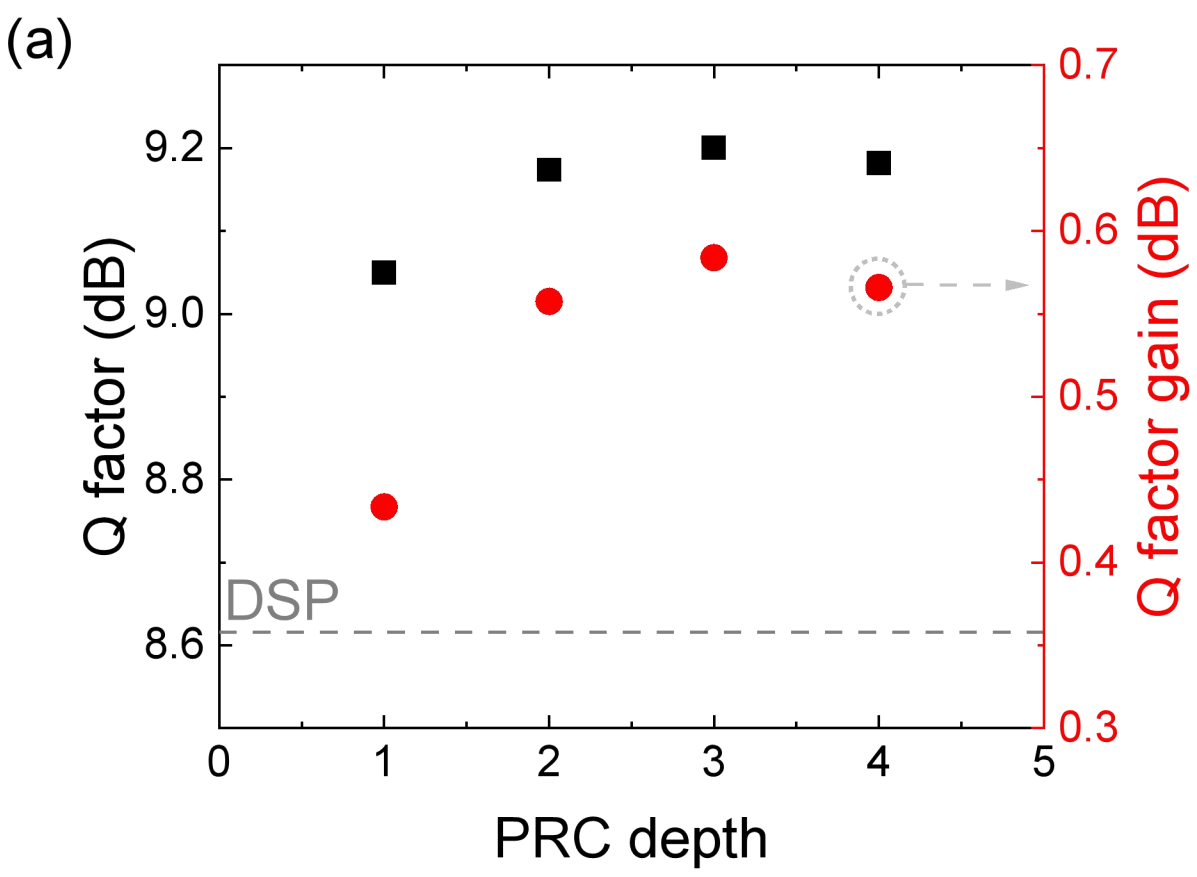

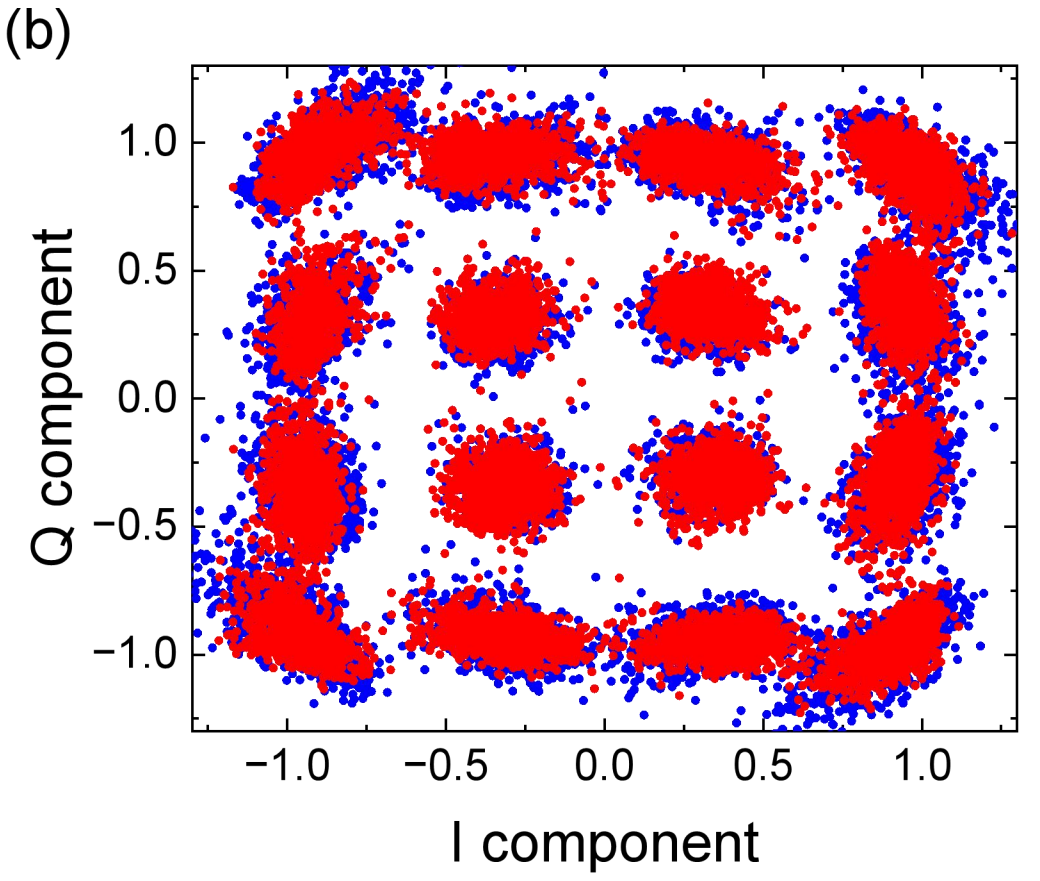

**FIG. 7.** (a) Impact of the PRC depth on the Q factor (squares) and the Q factor gain (dots). The dashed line is the Q factor (8.62 dB) of the DSP alone. (b) Constellation diagrams with (red) and without (blue) the nonlinear equalization by the PRC with three hidden layers. The Q factor gain is 0.58 dB, corresponding to a BER reduction of 44%.

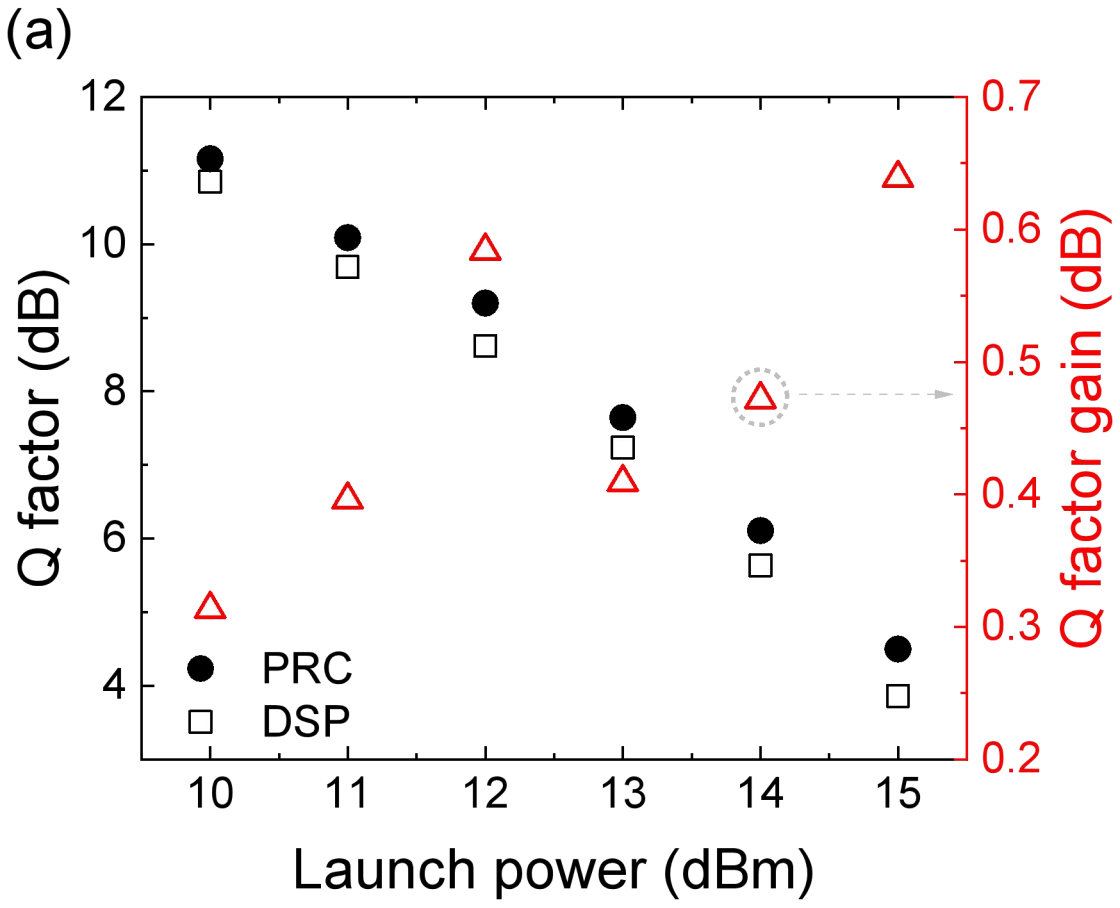

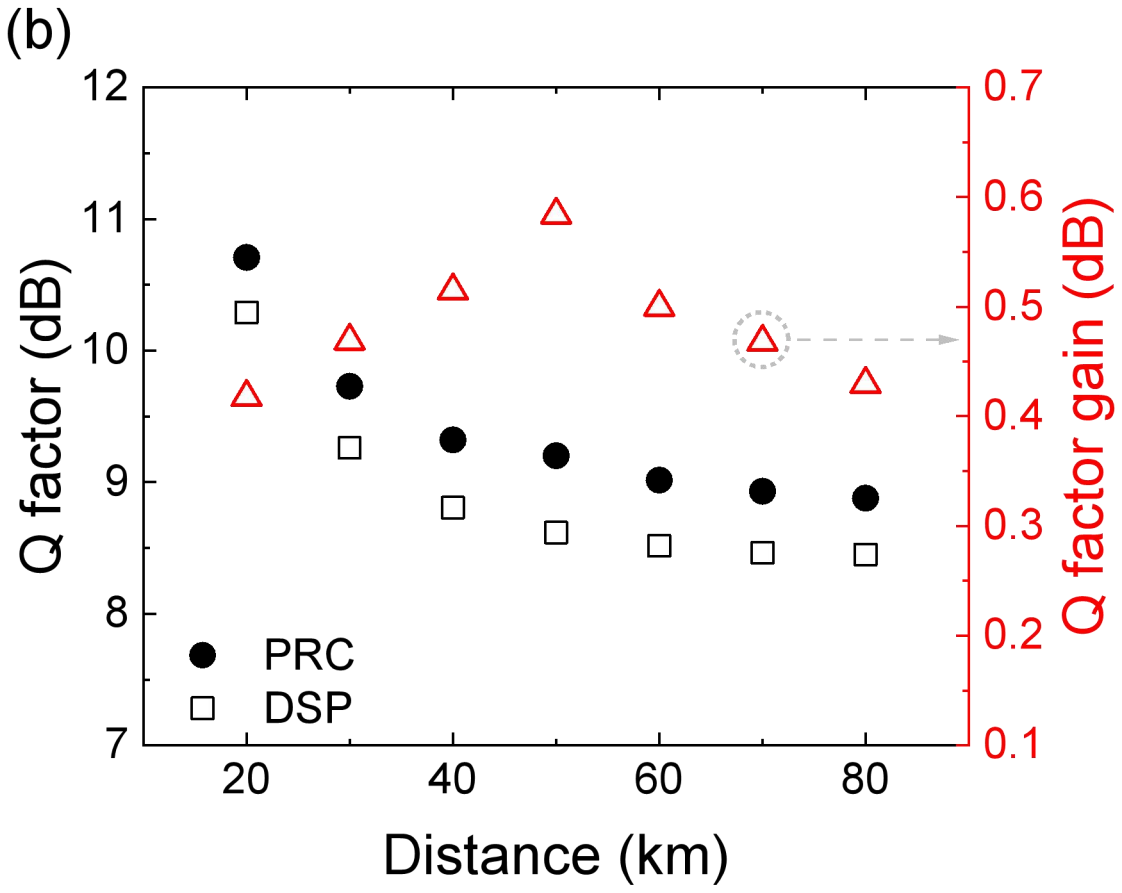

**FIG. 8.** Q factor (dots) and Q factor gain (triangles) of the deep PRC (a) versus the launch power and (b) versus the transmission distance. The depth of the PRC is 3, and the neuron number is 40 in each layer. The other operation parameters are listed in Table 1.

PRC decrease with increasing transmission distance as well, because of enhanced fiber nonlinearity accumulation and lower SNR. Aligned with the expectation, the maximum Q factor gain is achieved at 50 km, since the operation parameters in Table 1 are optimized at this distance. The gain drops down to 0.41 dB at 20 km at the shorter distance side, and down to 0.43 dB at 80 km at the longer distance side. Future work will optimize the operation parameters for every launch power and every transmission distance so as to achieve the best equalization performance.

## IV. DISCUSSION

In the experimental setup of Fig. 3, the I component and the Q component are sent to a single modulator in sequence one by one. However, in practice both components are output from the DSP module in parallel. In order to process the I component and Q component simultaneously, we propose the parallel architecture of the deep PRC as illustrated in Fig. 9. The deep PRC consists of two channels, where one channel processes the I component and the other channel processes the Q component, simultaneously. Each channel has a depth of two, although Fig. 7 proves that the optimal depth is three. This depth is limited by the 4 radio-frequency channels of the OSC used in the experiment, whereas a depth of three in combination with two PRC channels requires 6 OSC channels. The operation principle of Fig. 9 is similar to that of Fig. 3. The equalization performance of this parallel setup is tested by the same dataset of the 16-QAM signals.

Figure 10 shows the equalization performance of the dual-channel PRC with two hidden reservoir layers (squares). It is shown that the Q factor gain first increases from 0.31 dB at the launch power of 10 dBm up to the maximum of 0.55 dB at 12 dBm. Then, the gain decreases down to 0.38 dB at 13 dBm, because the operation parameters of the PRC are optimized at 12 dBm. The gain re-increases with rising launch power to 0.60 dB at 15 dBm. Generally, the Q factor gain of the dual-channel PRC with 2 hidden layers is similar to the single-channel one with 2 hidden layers (dots), while smaller than that of the single-channel one with 3 hidden layers (triangles). Therefore, the channels of the deep PRC can be easily scalable to four channels or more so as to deal with both polarizations of the 16-QAM signal, simultaneously.

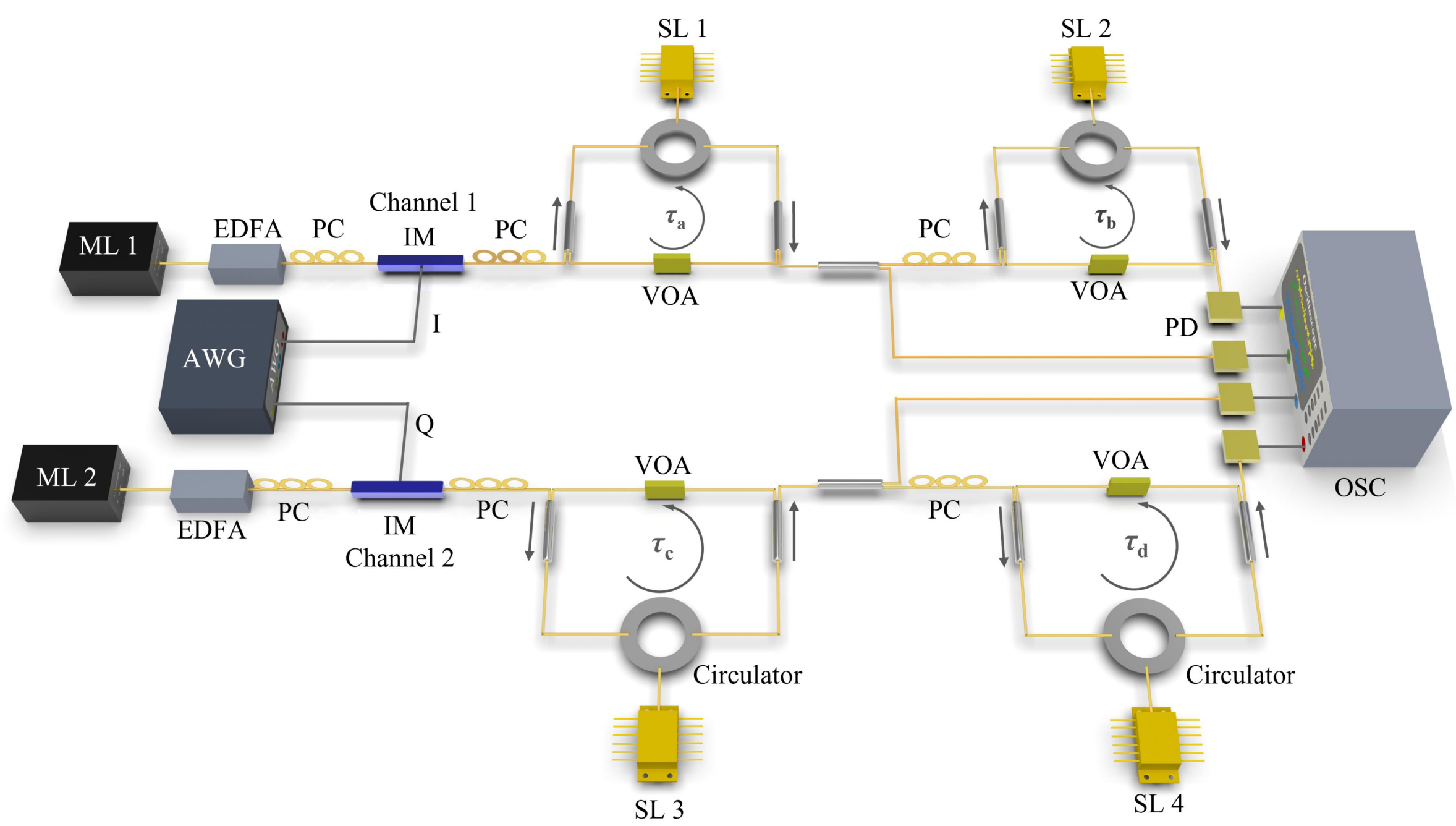

**FIG. 9.** Experimental setup of dual channel PRC with a depth of two. ML: master laser; EDFA: erbium-doped fiber amplifier; IM: intensity modulator; SL: slave laser; AWG: arbitrary waveform generator; OSC: oscilloscope; PD: photodiode; PC: polarization controller; VOA: variable optical attenuator.

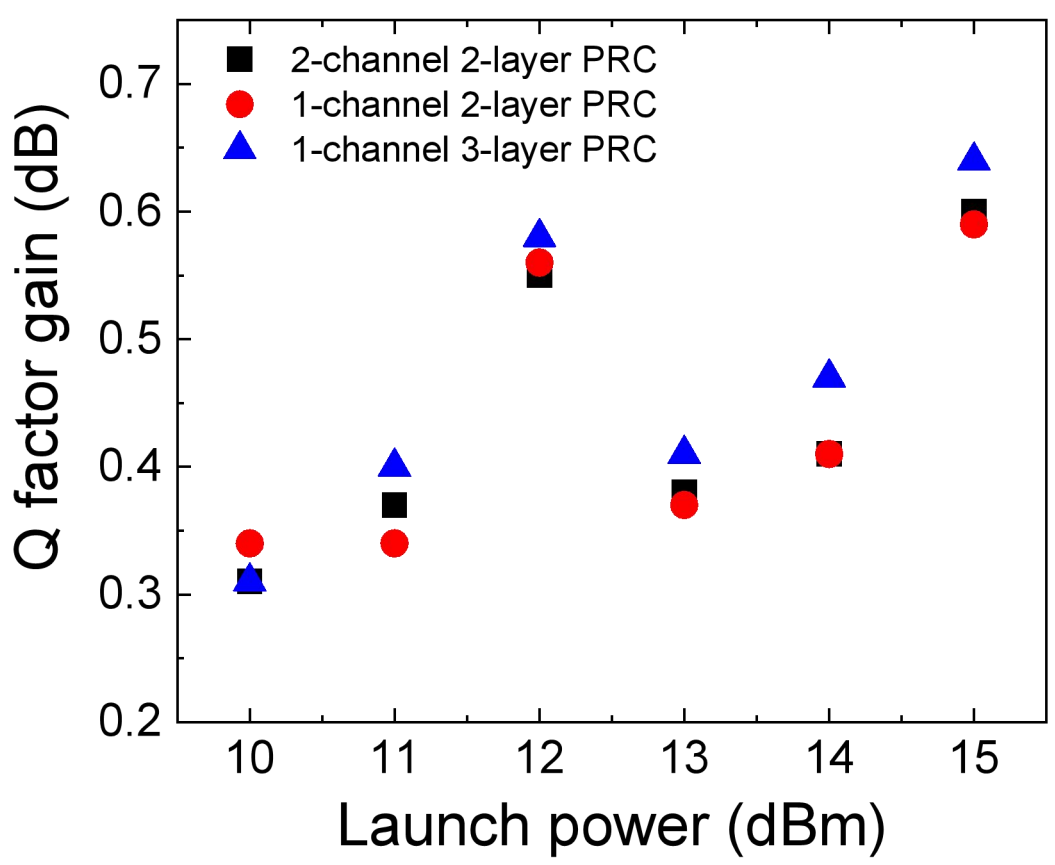

**FIG. 10.** Q factor gain of the deep PRC versus the launch power of the transmitter for dual-channel PRC with two layers (squares), the single-channel PRCs with two layers (dots) and three layers (triangles).

## V. CONCLUSION

In conclusion, we experimentally demonstrated the nonlinear equalization of 16-QAM signals using the deep PRC. The hidden reservoir layer of the deep PRC consists of a semiconductor laser with an optical feedback loop. Different hidden layers are connected in series through the cascading injection-locking technique. It is proved that the deep PRC shows strong capability in mitigating fiber nonlinear impairment. For the equalization of the 16-QAM signal of 240 Gbps with a launch power of 12 dBm and with a transmission distance of 50 km, the single-channel PRC with a depth of 3 achieves a Q factor gain of 0.58 dB (BER reduction by 44%). In addition, the dual-channel PRC with a depth of 2 processes the I component and the Q component of the 16-QAM signal in parallel, and achieves a Q factor gain of 0.55 dB (BER reduction by 42%).

## ACKNOWLEDGMENTS

This work was funded by Science and Technology Commission of Shanghai Municipality (24JD1402400, 24TS1401500), National Natural Science Foundation of China (62475152, 62475150) and ShanghaiTech AI Initiative (AI2026A06)

## AUTHOR DECLARATIONS

### Conflict of Interest

The authors have no conflicts to disclose.

## REFERENCES

[1]J. G. Proakis and M. Salehi, *Digital Communication* (McGraw-Hill, 2008).

[2]Z. Niu, H. Yang, L. Li, M. Shi, G. Xu, W. Hu, and L. Yi, "Learnable digital signal processing: A new benchmark of linearity compensation for optical fiber communications," Light: Sci. Appl. **13**, 188 (2024).
[3]E. Ip and J. M. Kahn, "Digital equalization of chromatic dispersion and polarization mode dispersion," J. Lightwave Technol. **25**, 2033–2043 (2007).
[4]E. Ip and J. M. Kahn, "Compensation of dispersion and nonlinear impairments using digital backpropagation," J. Lightwave Technol. **26**, 3416–3425 (2008).
[5]Z. Tao, L. Dou, W. Yan, L. Li, T. Hoshida, and J. C. Rasmussen, "Multiplier-free intrachannel nonlinearity compensating algorithm operating at symbol rate," J. Lightwave Technol. **29**, 2570–2576 (2011).
[6]R. D. Nowak and B. D. Van Veen, "Volterra filter equalization: A fixed point approach," IEEE Trans. Signal Process. **45**, 377–388 (1997).
[7]Q. Fan, G. Zhou, T. Gui, C. Lu, and A. P. T. Lau, "Advancing theoretical understanding and practical performance of signal processing for nonlinear optical communications through machine learning," Nat. Commun. **11**, 3694 (2020).
[8]P. J. Freire, E. Manuylovich, J. E. Prilepsky, and S. K. Turitsyn, "Artificial neural networks for photonic applications—From algorithms to implementation: Tutorial," Adv. Opt. Photonics **15**, 739–834 (2023).
[9]S. Zhang, F. Yaman, K. Nakamura, T. Inoue, V. Kamalov, L. Jovanovski, V. Vusirikala, E. Mateo, Y. Inada, and T. Wang, "Field and lab experimental demonstration of nonlinear impairment compensation using neural networks," Nat. Commun. **10**, 3033 (2019).
[10]L. Huang, Y. Xu, W. Jiang, L. Xue, W. Hu, and L. Yi, "Performance and complexity analysis of conventional and deep learning equalizers for the high-speed IMDD PON," J. Lightwave Technol. **40**, 4528–4538 (2022).
[11]Y. Liu, V. Sanchez, P. J. Freire, J. E. Prilepsky, M. J. Koshkouei, and M. D. Higgins, "Attention-aided partial bidirectional RNN-based nonlinear equalizer in coherent optical systems," Opt. Express **30**, 32908–32923 (2022).
[12]L. Yi, T. Liao, L. Huang, L. Xue, P. Li, and W. Hu, "Machine learning for 100 Gb/s/λ passive optical network," J. Lightwave Technol. **37**, 1621–1630 (2019).
[13]A. Argyris, J. Bueno, and I. Fischer, "Photonic machine learning implementation for signal recovery in optical communications," Sci. Rep. **8**, 8487 (2018).
[14]J. Vatin, D. Rontani, and M. Sciamanna, "Experimental realization of dual task processing with a photonic reservoir computer," APL Photonics **5**, 086105 (2020).
[15]B. Wang, Q. Xiao, T. Xu, L. Fan, S. Liu, Q. Kong, J. Dong, J. Zhang, and C. Huang, "An all-optical signal processor enabling terabit-per-second real-time equalization," Science **392**, eady5344 (2026).
[16]R. Van Assche, S. Masaad, E. Gooskens, S. Sackesyn, J. Van Kerrebrouck, X. Yin, and P. Bienstman, "Real-time optical signal equalization with a silicon photonic spatially distributed reservoir computer," Nat. Photonics (2026).
[17]S. Masaad, E. Gooskens, S. Sackesyn, J. Dambre, and P. Bienstman, "Photonic reservoir computing for nonlinear equalization of 64-QAM signals with a Kramers–Kronig receiver," Nanophotonics **12**, 925–935 (2023).
[18]A. Zelaci, S. Masaad, and P. Bienstman, "Reservoir computing for equalization in a self-coherent receiver scheme," Opt. Express **32**, 40326–40339 (2024).
[19]Y. Zhang, T. Liu, J. Zhao, and T. Xu, "Delay-based reservoir computing for signal recovery in optical communication systems," Opt. Fiber Technol. **95**, 104395 (2025).
[20]S. Masaad, S. Sackesyn, S. Sygletos, and P. Bienstman, "Experimental demonstration of 4-port photonic reservoir computing for equalization of 4 and 16 QAM signals," J. Lightwave Technol. **42**, 8555–8563 (2024).
[21]D. Brunner, M. C. Soriano, C. R. Mirasso, and I. Fischer, "Parallel photonic information processing at gigabyte per second data rates using transient states," Nat. Commun. **4**, 1364 (2013).
[22]K. Vandoorne, P. Mechet, T. Van Vaerenbergh, M. Fiers, G. Morthier, D. Verstraeten, B. Schrauwen, J. Dambre, and P. Bienstman, "Experimental demonstration of reservoir computing on a silicon photonics chip," Nat. Commun. **5**, 3541 (2014).
[23]L. Larger, M. C. Soriano, D. Brunner, L. Appeltant, J. M. Gutiérrez, L. Pesquera, C. R. Mirasso, and I. Fischer, "Photonic information processing beyond Turing: An optoelectronic implementation of reservoir computing," Opt. Express **20**, 3241–3249 (2012).
[24]G. Donati, C. R. Mirasso, M. Mancinelli, L. Pavesi, and A. Argyris, "Microring resonators with external optical feedback for time-delay reservoir computing," Opt. Express **30**, 522–537 (2022).
[25]G. P. Agrawal, *Nonlinear Fiber Optics* (Springer, 2001).
[26]X. Liu, A. R. Chraplyvy, P. J. Winzer, R. W. Tkach, and S. Chandrasekhar, "Phase-conjugated twin waves for communication beyond the Kerr nonlinearity limit," Nat. Photonics **7**, 560–568 (2013).
[27]C. Gallicchio, A. Micheli, and L. Pedrelli, "Deep reservoir computing: A critical experimental analysis," Neurocomputing **268**, 87–99 (2017).
[28]C. Gallicchio, A. Micheli, and L. Pedrelli, "Design of deep echo state networks," Neural Networks **108**, 33–47 (2018).
[29]B.-D. Lin, Y.-W. Shen, J.-Y. Tang, J. Yu, X. He, and C. Wang, "Deep time-delay reservoir computing with cascading injection-locked lasers," IEEE J. Sel. Top. Quantum Electron. **29**, 7600408 (2023).
[30]Y.-W. Shen, R.-Q. Li, G.-T. Liu, J. Yu, X. He, L. Yi, and C. Wang, "Deep photonic reservoir computing recurrent network," Optica **10**, 1745–1751 (2023).
[31]L. Appeltant, M. C. Soriano, G. Van der Sande, J. Danckaert, S. Massar, J. Dambre, B. Schrauwen, C. R. Mirasso, and I. Fischer, "Information processing using a single dynamical node as complex system," Nat. Commun. **2**, 468 (2011).
[32]J.-Y. Tang, B.-D. Lin, Y.-W. Shen, R.-Q. Li, J. Yu, X. He, and C. Wang, "Asynchronous photonic time-delay reservoir computing," Opt. Express **31**, 2456–2466 (2023).
[33]R.-Q. Li, Y.-W. Shen, Z. Niu, G. Xu, J. Yu, X. He, L. Yi, and C. Wang, "Deep photonic reservoir computer for nonlinear equalization of 16-level quadrature amplitude modulation signals," APL Mach. Learn. **3**, 026113 (2025).
[34]G. Goldfarb and G. Li, "Chromatic dispersion compensation using digital IIR filtering with coherent detection," IEEE Photonics Technol. Lett. **19**, 969–971 (2007).
[35]A. Leven, N. Kaneda, U. V. Koc, and Y. K. Chen, "Frequency estimation in intradyne reception," IEEE Photonics Technol. Lett. **19**, 366–368 (2007).
[36]M. G. Taylor, "Phase estimation methods for optical coherent detection using digital signal processing," J. Lightwave Technol. **27**, 901–914 (2009).
[37]L. B. Du, D. Rafique, A. Napoli, B. Spinnler, A. D. Ellis, M. Kuschnerov, and A. J. Lowery, "Digital fiber nonlinearity compensation: Toward 1-Tb/s transport," IEEE Signal Process. Mag. **31**, 46–56 (2014).
[38]O. Vassilieva, I. Kim, and T. Ikeuchi, "Enabling technologies for fiber nonlinearity mitigation in high-capacity transmission systems," J. Lightwave Technol. **37**, 50–60 (2019).
[39]Z. Niu, H. Yang, L. Li, M. Shi, C. Zeng, C. Dai, G. Xu, H. Zhao, and L. Yi, "Intelligent fiber transmission simulation," https://ifibertrans.sjtu.edu.cn/ (2022).
[40]J. Ohtsubo, *Semiconductor Lasers: Stability, Instability and Chaos* (Springer, 2013).
[41]Y. Kuriki, J. Nakayama, K. Takano, and A. Uchida, "Impact of input mask signals on delay-based photonic reservoir computing with semiconductor lasers," Opt. Express **26**, 5777–5788 (2018).
[42]T. Hülser, F. Köster, L. Jaurigue, and K. Lüdge, "Role of delay times in delay-based photonic reservoir computing," Opt. Mater. Express **12**, 1214–1231 (2022).
[43]W. Freude, R. Schmogrow, B. Nebendahl, M. Winter, A. Josten, D. Hillerkuss, S. Koenig, J. Meyer, M. Dreschmann, M. Huebner, C. Koos, J. Becker, and J. Leuthold, "Quality metrics for optical signals: Eye diagram, Q-factor, OSNR, EVM and BER," in *2012 14th International Conference on Transparent Optical Networks (ICTON)* (IEEE, 2012), pp. 1–4.
[44]C. Gallicchio and A. Micheli, "Architectural richness in deep reservoir computing," Neural Comput. Appl. **35**, 24525–24542 (2023).
[45]J. L. Wei, J. D. Ingham, D. G. Cunningham, R. V. Penty, and I. H. White, "Performance and power dissipation comparisons between 28 Gb/s NRZ, PAM, CAP and optical OFDM systems for data communication applications," J. Lightwave Technol. **30**, 3273–3280 (2012).
[46]G. P. Agrawal, *Fiber-Optic Communication Systems* (John Wiley & Sons 2012).